\documentclass[prb,aps,twocolumn,superscriptaddress]{revtex4-2}

\usepackage{color}
\usepackage{amssymb}
\usepackage{amsmath}
\usepackage{graphicx}
\usepackage{xcolor}
\usepackage[normalem]{ulem}
\usepackage[unicode=true,pdfusetitle,hypertexnames=false,
 bookmarks=true,bookmarksnumbered=false,bookmarksopen=false,
  breaklinks=true,pdfborder={0 0 0},pdfborderstyle={},backref=false,colorlinks=true]
   {hyperref}
\usepackage{subcaption}

\definecolor{darkgreen}{RGB}{89, 175, 99}
\definecolor{grey}{rgb}{.6,.6,.6}
\definecolor{eggplant}{RGB}{180,33,147}

\newcommand{\ket}[1]{| #1 \rangle}
\newcommand{\bra}[1]{  \langle #1 |}

\usepackage{cancel}
\usepackage{layouts}

\begin{document}

\title{Feasibility of performing quantum chemistry calculations on quantum computers}

\author{Thibaud Louvet}
\affiliation{
PHELIQS, Université  Grenoble Alpes, CEA, Grenoble INP, IRIG, Grenoble 38000, France
}

\author{Thomas Ayral}
\affiliation{
Eviden Quantum Laboratory, Les Clayes-sous-Bois, France
}

\author{Xavier Waintal}
\affiliation{
PHELIQS, Université Grenoble Alpes, CEA, Grenoble INP, IRIG, Grenoble 38000, France
}
\date{\today}

\begin{abstract}
Quantum chemistry is envisioned as an early and disruptive application for quantum computers.
Yet, closer scrutiny of the proposed algorithms shows that there are considerable difficulties along the way.
Here, we propose two criteria for evaluating two leading quantum approaches for finding the ground state of molecules.
The first criterion applies to the variational quantum eigensolver (VQE) algorithm.
It sets an upper bound to the level of  imprecision/decoherence that can be tolerated in quantum hardware as a function of the targeted precision, the number of gates and the typical energy contribution from states populated by decoherence processes.
We find that decoherence is highly detrimental to the accuracy of VQE and 
performing relevant chemistry calculations would require performances that are
expected for fault-tolerant quantum computers,
not mere noisy hardware, even with advanced error mitigation techniques.
Physically, the sensitivity of VQE to decoherence originates from the
fact that, in VQE, the spectrum of the studied molecule has no correlation with the
spectrum of the quantum hardware used to perform the computation.

The second criterion applies to the quantum phase estimation (QPE) algorithm, which is often presented as the go-to replacement of VQE upon availability of (noiseless) fault-tolerant quantum computers.
QPE requires an input state with a large enough overlap with the sought-after ground state.
We provide a criterion to estimate quantitatively this overlap based on the energy and the energy variance of said input state.
Using input states from a variety of state-of-the-art classical methods, we show that the scaling of this overlap with system size does display the standard orthogonality catastrophe, namely an exponential suppression with system size. This in turns leads to an exponentially reduced QPE success probability. 
\end{abstract}

\maketitle

There exists a hierarchy in the applications that have been proposed for quantum computers, from Shor's algorithm \cite{Shor1994} (exponentially faster than its known classical counterparts) to Grover's algorithm \cite{Grover1997} (up to quadratically faster than its classical counterparts---but less specialized, and whose practical relevance has been questioned, see \cite{Stoudenmire2023}
and references therein), to near-term algorithms such as the variational quantum eigensolver (VQE) \cite{Peruzzo2014}, which have no a priori parametric advantage over their classical counterparts, but might have a practical one \cite{Chan2024}.
This hierarchy ends with quantum simulations, also dubbed "analog quantum computing", where one trades the quantum gate model for a direct manipulation of the hardware Hamiltonian. 
Going down this hierarchy, one gives up on the advantage provided by the quantum computer in terms of the degree of generality of the applications and arguably the expected provable speedup.
In turn, the requirements on the hardware become less drastic, and the hope for speedup stems from the idea that quantum processors, as medium-scale interacting many-body systems, could capture important physics of the quantum many-body problems encountered in many fields \cite{Ayral2023b}.

This article focuses on one many-body problem that has been put forward as a possible near-term application for quantum computers: quantum chemistry \cite{Aspuru-Guzik2005,McArdle2018a, Bauer2020,Tilly2022}, and in particular the search for the ground state energy of molecules.
The standard algorithms are VQE \cite{Kuhn2019,Elfving2020,Cerezo2021,Gonthier2022,Tilly2022} for noisy, intermediate scale quantum (NISQ) processors, or its fault-tolerant counterpart, the quantum phase estimation (QPE) algorithm \cite{Wecker2015,Vonburg2021,Beverland2022}.
Despite great expectations, it is a very difficult exercise to extrapolate the existing hardware capabilities to estimate whether a quantum advantage will be eventually reached.
Here we take a somewhat reverse approach and derive {\it necessary} conditions to obtain such an advantage, thereby defining constraints that the hardware must fulfill if an advantage is to be obtained.
We discuss, in turn, VQE and QPE.

To define the notion of quantum advantage, one must know what one is competing with, i.e. the current state of the art of numerical quantum chemistry methods.
The entrance cost for a quantum computer in this field is very stiff: existing numerical techniques are in fact quite good despite the apparent exponential complexity of the problem.
Consider a problem of $N$ electrons distributed among $M$ orbitals (hence at least as many qubits, assuming inactive orbitals have been dealt with thanks to classical methods).
A typical quantum chemistry Hamiltonian takes the form
\begin{equation}
	H = \sum_{ij} h_{ij} c^\dagger_i c_j + \frac{1}{2} \sum_{ijkl} v_{ijkl} c^\dagger_i c^\dagger_j c_k c_l
	\label{eq:hamiltonian}
\end{equation}
where $c^\dagger_i$ (respectively $c_i$) creates (respectively destroys) an electron on the spin-orbital $i\in \{1,...,M\}$,  $h_{ij}$ corresponds to the kinetic and nuclei Hamiltonian matrix elements and $v_{ijkl}$ corresponds to the electron-electron Coulomb matrix elements. The goal of VQE and QPE
is to find the ground state $|\Psi_0\rangle$ of $H$ and in particular calculate the associated energy $E_0$.

Such a problem can be addressed by a large spectrum of classical techniques with different precisions and computational costs.
Mean-field techniques (improved with e.g. GW diagrammatic approaches) can reach a precision of $\approx 10$~ mHa for up to $M > 5000$ and hundreds of electrons
(see e.g. \cite{Bruneval2021} for typical calculations), which is sufficient for calculating dissociation energies (including phonon spectra), the energy liberated by a chemical reaction, or ionization energies.
Reaching an accuracy of ${\approx~1 \mathrm{~mHa}~\approx~300\mathrm{~K}}$, needed to compute activation energies, requires more precise techniques such as coupled cluster techniques, whose computational complexity scales as $N^2 (M-N)^4$ (up to two electron-hole excitations), $N^3 (M-N)^5$ (three excitations) or more, making these approaches impractical beyond a few hundred orbitals \cite{Bartlett2007}.
The above approaches work well for many molecules.
However they typically fail for \emph{strongly correlated} molecules (e.g. containing transition metals), which require even more complex methods such as the density-matrix renormalization group method (DMRG, \cite{Chan2011}).
These methods can typically handle only tens of orbitals.
It is this latter case that has been identified as the early target for quantum algorithms, in particular in the context of nitrogen fixation for fertilizers \cite{Reiher2017}, although addressing this problem might not be as ground breaking as initially surmised \cite{Chan2024}.

Before introducing the two criteria, let us emphasize that the size $M$ of the basis set plays a very important role in the very definition of the quality of a ground-state energy estimation, as pointed out in Ref.~\cite{Elfving2020}.
One indeed needs to distinguish precision---the ability to correctly solve the many-body problem for given $N$ and $M$---and accuracy---the ability to reach large enough $M$ to approach the complete basis set limit $M\rightarrow\infty$ for a given $N$.
For instance, the "spherical cow" of the VQE algorithm, namely the benchmark of the H$_2$ energy versus H$-$H distance, almost always uses only a single orbital per hydrogen atom (STO-3G or STO-6G basis set) and corresponds to an accuracy of  at best $28$ mHa, even if one reaches perfect precision. 
The true target of quantum computers is chemical accuracy ($< 1.6$ mHa), which always means using much larger basis sets.
This means more qubits and, as we shall see, more severe requirements on the hardware.

\section{A criterion for the variational quantum eigensolver (VQE): maximum hardware error}

We start our analysis with VQE, which has been proposed as a possible algorithm that could be usable with noisy quantum computers. 
VQE suffers from three important difficulties: (I) its sensitivity to statistical uncertainty,
(II) the difficulty to perform the optimization of the wavefunction and (III) its sensitivity to
hardware imprecisions and decoherence. In this section we briefly review the VQE algorithm as well as problems I and II, then proceed with an in-depth discussion of problem III. We conclude with a point-by-point comparison of VQE and its natural classical counterpart, the variational Monte Carlo (VMC) algorithm.

\subsection{Description of VQE and its challenges}
VQE is a variational algorithm that consists in preparing a parameterized wavefunction
\begin{equation}
\ket{\Psi_V(\vec\theta)} = U(\vec\theta)\ket{0},
\end{equation}
(with $N_\theta$ variables $\vec\theta = (\theta_1,\theta_2...\theta_{N_\theta})$), with a quantum computer and minimizing its energy
\begin{equation}
E_V(\vec\theta) = \bra{\Psi_V} H \ket{\Psi_V}
\end{equation}
with classical optimization algorithms.
The use of a quantum computer to prepare the variational wavefunction allows to consider classes of wavefunctions that could a priori not be accessed using classical computers.
Starting from a given problem, one constructs the corresponding Hamiltonian representation, choosing a given molecular basis (e.g., Gaussian basis functions) and representation  (e.g., second quantization).
Then, the Hamiltonian is encoded by e.g., a Jordan-Wigner transformation into a qubit representation.
One key point of VQE is the choice of an ansatz for the variational wavefunction, namely the parameterized quantum circuit $U(\vec\theta)$.
The ansatz must be "expressive" enough so that it covers the interesting part of the Hilbert space, while using as few parameters as possible, and allowing for a shallow circuit representation due to the limited circuit depths and run times available in NISQ, real-life experiments.

VQE has a number of well-documented shortcomings.
One important drawback is the large number of individual measurements required to estimate the expectation value of the energy $E_V$ (problem I).
This is due to the probabilistic nature of measurements intrinsic to quantum mechanics.
Indeed, to measure $E_V$, the Hamiltonian is decomposed as a sum of groups of mutually-commuting Pauli strings:
\begin{equation}
H = \sum_{\alpha =1}^{N_\alpha} H_\alpha
\end{equation}
with $H_\alpha = \sum_{k_{\alpha}} \lambda_{k_{\alpha}} P_{k_{\alpha}}$,
with $P_{k_\alpha}$ of the form $\sigma_1 \otimes \cdots \otimes \sigma_n$ (where $\sigma_i$ is one of the Pauli matrices acting on qubit $i$ ($X$, $Y$, $Z$ or identity)).
The expectation value of each term $H_\alpha$ is measured separately: for each $\alpha$, one repeatedly prepares state $\ket{\Psi_V}$ and measures each $P_{k_\alpha}$ in the common eigenbasis of the $P_{k_\alpha}$'s.
The outcomes are averaged to get an estimate of $\bra{\Psi_V} P_{k_\alpha}\ket{\Psi_V}$. 
The statistical error stemming from the finite number $N_s$ of repetitions, or shots, is of the form $\epsilon \approx \sigma (P_{k_\alpha}) / \sqrt{N_s}$ where $\sigma(P_{k_\alpha})$ is the standard deviation of individual measurements.
Therefore, the run time of computing $E_V$, which is proportional to $N_s$, scales as $1/\epsilon^2$.
This scaling is a direct consequence of the fact that energy measurements in VQE are based on a classical averaging of individual measurement outcomes.
Even more advanced methods like classical shadows \cite{Huang2020} also suffer from this scaling, although they can alleviate the dependence of the run time on the number of terms to be measured.

This unfavorable scaling of VQE leads to quite large run time estimates if good enough accuracies $\epsilon$ are to be reached: in 2015, a study estimated that to evaluate the ground state energy of Fe$_2$Se$_2$ in the minimal STO$-3$g orbital basis with chemical precision would require a prohibitive number of $10^{19}$ samples~\cite{Wecker2015}.
Another study estimated the runtime to estimate the combustion energy for methane to be of 1.9 days and 71 days for ethanol, finding no practical advantage compared to state-of-the-art classical techniques (see Table II in \cite{Gonthier2022}).
In a recent review, combining the state of the art methods in each part of the VQE pipeline and assuming a gate time of $100$~ns, the total runtime for one iteration of VQE on the Cr$_2$ molecule is estimated at 25 days. Taking into account the number of iterations required to reach the minimum in parameter space yields an estimate of 24 years~\cite{Tilly2022}.

This leads us to another major issue of VQE, namely the convergence of the optimization towards an energy minimum (problem II).
There is no general guarantee that VQE converges. Like other variational techniques, it relies on heuristics.
In fact, the variance of the energy landscape $E_V(\vec\theta)$ over the parameter space happens to be exponentially small in many settings, a phenomenon dubbed the barren plateau phenomenon \cite{McClean2018}.
One such setting is when using very expressive circuits \cite{Ragone2023, Larocca2024}: these circuits generically lead to a exponentially vanishing variance and thus exponential run times as the optimization gets stuck on plateaus.
This average statement does not exclude finding initial parameters that escape those plateaus, but finding these parameters will likely involve heavy classical preprocessing, or adaptive ansatz constructions like ADAPT-VQE \cite{Grimsley2019}.

In the following, we will place ourselves in the optimistic situation where the available number of shots is large enough to reach chemical accuracy---thus getting rid of the statistical uncertainty of VQE---and where we have escaped the barren plateau issue. 
We shall thus only be concerned with the effect of decoherence on VQE.

\subsection{The decoherence versus precision criterion}

In this section, we disregard statistical uncertainty and barren plateaus, and aim at answering a more pragmatic short-term question: 
what is the level of decoherence sustainable to attain a targeted precision on the ground state energy? (problem III). Note that we use the terminology "decoherence" generically here to account for all the imperfections of the quantum hardware including actual decoherence, but also more mundane limitations to the precision such as the effect of crosstalk, $1/f$ noise, imprecise pulse sequence \cite{Waintal2024}.

\subsubsection{Derivation of the criterion}

We measure the effect of noise on the hardware with the fidelity $F=\bra{\Psi_V}  \rho \ket{\Psi_V}$.
It expresses how the density matrix $\rho$ of the quantum computer after a noisy execution of the variational circuit $U(\vec{\theta})$ differs from the expected one, $\ket{\Psi_V} \bra{\Psi_V}$.
$F<1$ implies that $\rho$ can always be written in the form
\begin{equation}
\rho = F \ket{\Psi_V} \bra{\Psi_V} + (1-F) \rho_{\rm noise},
\end{equation}
where $\rho_{\rm noise}$ is the part of the density matrix that results from decoherence.
The resulting energy is given by $E = \mathrm{Tr}(\rho H) =  E_V + \Delta E$ with a noise-induced error $\Delta E$ defined as
\begin{equation}
\label{eq:error}
\Delta E = (1-F) [E_{\rm noise} - E_V],
\end{equation}
with $E_{\rm noise} = {\rm Tr } (\rho_{\rm noise} H)$.
A large corpus of experiments and theory \cite{Cai2022, Ayral2023}, including the seminal "quantum supremacy" experiment by Google \cite{Arute2019}, shows that the fidelity decays exponentially with the total number of applied gates $N_g$,
\begin{equation}
\label{eq:fid}
F \approx e^{-\epsilon N_g}
\end{equation}
where $\epsilon$ is the average error per gate. This accumulation of errors puts a severe constraint on the number of gates and thereby on the performance of noisy quantum computers~\cite{Takagi2021,Aharonov2023}.  
The fact that different errors get combined in a multiplicative way can be put on firm mathematical basis only for certain error models, but is widely expected to hold more generally.

Suppose that one targets a precision  $\eta_{\rm chem}$ in the calculation with 
$\eta_{\rm chem} \ll E_\mathrm{noise} - E_V$. It follows that $F$ must be very close to unity. 
Combining the above equations, imposing $\Delta E\le \eta_{\rm chem}$ leads to
$\epsilon < \epsilon_{\rm max}$ with 
\begin{equation}
\label{eq:vqe_criterion}
\epsilon_{\rm max} =  \frac{\eta_{\rm chem}}{(E_{\rm noise}-E_V) N_g}.
\end{equation}
This is our quantitative criterion for using VQE on a given hardware, molecule and ansatz.
Let us emphasize that, in practice,  $N_g$  depends implicitly on $\eta_{\rm chem}$: good precision requires a complex ansatz, and hence a large number of gates.
Note that $\epsilon_{\rm max}$ is the effective error that one obtains \emph{after} one has done one's best to decrease the error level, including any quantum error correction scheme and/or error mitigation. Alternatively, one can write the same criterion in terms of the \emph{bare} error level
$\epsilon_{\rm bare}$, i.e. the error level as measured in randomized benchmarking on single qubits and pairs of qubits.
In the leading quantum hardware, the average error per gate is dominated by the two-qubit gates for which $\epsilon_{\rm bare} \le 1\%$.
 If one uses (possibly costly) error mitigation techniques that reduces the error in post processing by a factor $1/G$, we can write the above estimate in terms of the bare error level by simply replacing,
 \begin{equation}
 1-F \rightarrow G (1-F) 
 \end{equation}
 which in the limit of small $\epsilon$ ($F$ close to unity) amounts effectively to using an 
 average error per gate given by $\epsilon = \epsilon_{\rm bare} G$.
 In terms of the bare error, the criterion~\eqref{eq:vqe_criterion} is therefore slightly less stringent. One arrives at $\epsilon_{\rm bare} < \epsilon_{\rm max, b}$ with
\begin{equation}
\label{eq:vqe_criterion2}
\epsilon_{\rm max, b} =  \frac{\eta_{\rm chem}}{(E_{\rm noise}-E_V) N_g G}.
\end{equation}
Values as small as $G=0.001-0.01$ have been reported in the literature for relatively small systems, but it is unclear if such large gains can be maintained when the system is scaled up owing to the generically exponential cost of error mitigation (as discussed in more detail in 
Sec.~\ref{sec:error_mitig}). Since our goal is to find constraints on the hardware and the value of $G$ is strongly hardware dependent, in the following, we focus on the criterion for the effective error level $\epsilon_{\rm max}$.

\begin{figure}
	\includegraphics[width=7cm]{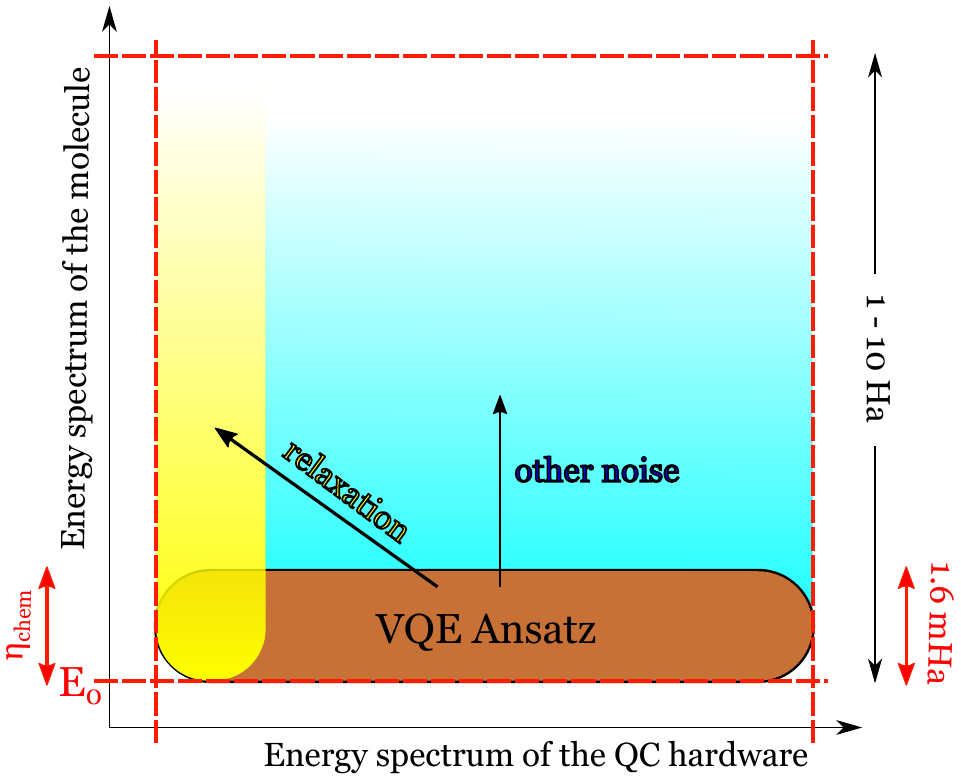}
	\caption{Schematic of the difference between the hardware spectrum and that of the studied molecule. In terms of the target eigenstates, the VQE ansatz is close to the ground state of the molecule; in terms of the hardware eigenstates, it is made of arbitrary, both low- and high-energy states.
	On the other hand, the hardware noise can populate arbitrarily high excited states of the studied molecule.
	For instance, relaxation can populate the hardware ground state, which consists of arbitrarily high energy states in terms of the target eigenstates. \label{fig:schema}}
\end{figure}

\begin{figure}
	\includegraphics[width=\columnwidth]{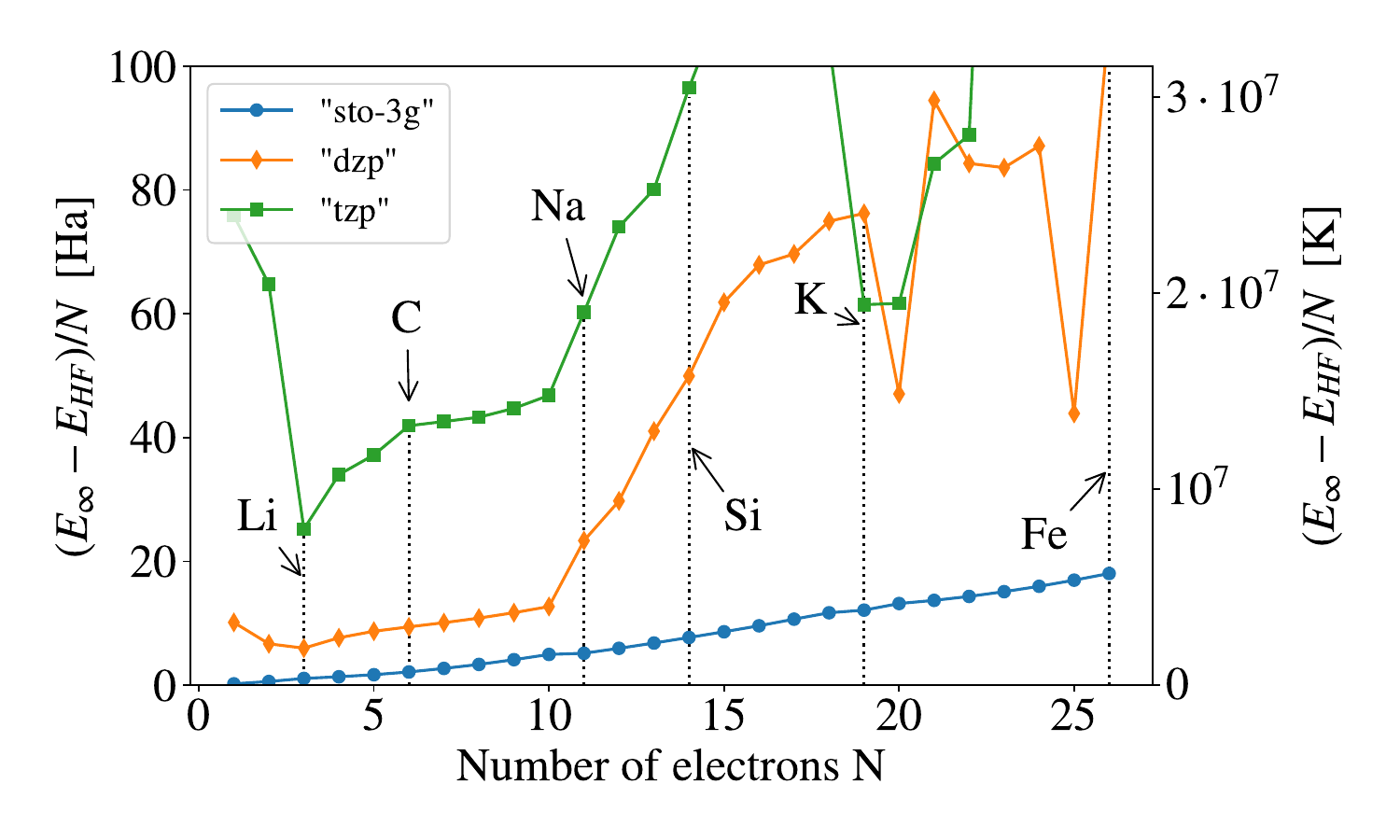}
	\caption{\label{fig:atoms} Difference between the energy of the infinite temperature state $E_\infty$ and the energy of the ground state, computed within the Hartree-Fock approximation $E_{HF}$, per electron, for the first atoms in the periodic table, from H to Fe for three common basis sets.}
\end{figure}

\subsubsection{Discussion: scaling of the $(E_\mathrm{noise} - E_V)$ energy scale}

We now argue that the energy scale in the denominator, $E_\mathrm{noise} - E_V$, is generically a large energy scale---of the order of the bare matrix elements of the Hamiltonian, Hartrees or tens of Hartrees---and even generically scales very unfavorably as the square of the number of electrons.

To understand the origin of this behavior, we need to remember that there are two very different Hamiltonians in the problem with two very different energy spectra: the target Hamiltonian $H$
(the one we want to find the ground state of) and the Hamiltonian $H^{\rm hw}$ that describes the quantum hardware (sometimes called the "resource Hamiltonian", see e.g., Ref \cite{Kokail2018}).
More generally, the dynamics of the quantum computer hardware is described by a time-dependent Lindbladian, but for the purpose of this discussion, it is sufficient to consider its 
Hamiltonian. We thus have two sets of eigenstates and eigenenergies,
\begin{equation}
H |\Psi_i\rangle = E_i |\Psi_i\rangle
\end{equation}
and
\begin{equation}
H^{\rm hw} |\Psi_i^{\rm hw}\rangle = E_i^{\rm hw} |\Psi^{\rm hw}_i\rangle
\end{equation}
Our main observation is that in general the target Hamiltonian $H$ is very different from the hardware Hamiltonian $H^{\rm hw}$. Indeed the same hardware Hamiltonian must be used for calculating different molecules so it generically has no correlations with the latter. This is a constraint that a \emph{gate-based} quantum circuit approach (as opposed to an analog quantum simulation) must obey in exchange for its universality. This situation is different from e.g annealing algorithms where the hardware Hamiltonian is supposed to be slowly deformed into the target Hamiltonian to end up in a low-energy state of the target Hamiltonian. This is also different from analog computing models, which strive to construct hardware Hamiltonians that are as close as possible to the target Hamiltonian, and may thus have similar low-energy subspaces. In other words, gate-based models, to gain universality, lose the physical constraints (for instance, symmetries) that could guarantee that the hardware Hamiltonian indeed shares properties with the target Hamiltonian.

It follows that the sought-after $\ket{\Psi_V}$ is generically composed of a superposition of
eigenstates of the hardware Hamiltonian that have arbitrary (small and large) energies:
\begin{equation}
\ket{\Psi_V} = \sum_i c_i |\Psi^{\rm hw}_i\rangle
\end{equation}
with $c_i$ having non zero components in the entire spectrum $E_i^{\rm hw}$ (see Fig.~\ref{fig:schema} for a schematic). Conversely the hardware eigenstates involve very 
high energies of the molecule (possibly even states with different number of electrons).
Why is this observation important? Because the mechanisms that lead to imprecisions and decoherence
usually have a simple structure related to the hardware spectrum. This in turn, means that these mechanisms will be structureless with respect to the target spectrum. A simple example is energy relaxation of the hardware---a ubiquitous process. Its  fixed point is the thermal density matrix $\rho \propto \sum_i e^{-E_i/kT} \ket{\Psi^{\rm hw}_i}\bra{\Psi^{\rm hw}_i}$.
 As another example, consider the admixture of a small amount $(1-F)$ of, say the ground state, $\ket{\Psi^{\rm hw}_0}\bra{\Psi^{\rm hw}_0}$ in the density matrix: it will lead to a noise energy 
 $E_{\rm noise} = \bra{\Psi^{\rm hw}_0}H \ket{\Psi^{\rm hw}_0}$. Since $\ket{\Psi^{\rm hw}_0}$
 is composed of a superposition of "target" states $\ket{\Psi_i}$ of arbitrary "target" energies $E_i$, it follows that
 $E_{\rm noise}$ typically lies in the middle of the spectrum of the target Hamiltonian,
\begin{equation}
E_{\rm noise} \approx E_{2^{n-1}}
\label{eq:noise}
\end{equation} 
We conjecture that Eq.\eqref{eq:noise} is generic for most, if not all, hardware errors and decoherence processes. It is a natural consequence of the versatility of gate-based quantum computers: the hardware errors are structureless in the target energy spectrum space. Supposing otherwise amounts
to adapting the hardware to the spectrum of the particular molecule that is being studied, which is precisely what a calculator is supposed \emph{not to} do (as opposed to a simulation).

A simple noise model where $E_{\rm noise}$ can be calculated explicitly is the (global) depolarizing channel, namely the map $\rho\rightarrow (1-\epsilon) \rho + \epsilon I_d/2^n$, where $I_d$ is the identity matrix. For this model, $E_{\rm noise} = E_\infty = \mathrm{Tr} (H) / 2^n$, where $E_\infty$ is the equilibrium energy of the target Hamiltonian $H$ at {\it infinite} temperature. 
We find once again that the (depolarizing) noise populates eigenstates of $H$ of {\it arbitrary} large energies. Local Pauli errors have the same fixed point \cite{Wang2021a} and would provide the same energy contribution $E_{\infty}$.

To continue, we will evaluate $E_{\rm noise}$ for some concrete cases. For definiteness, we will 
focus on $E_{\rm noise}= E_{\infty}$, but $E_{\rm noise}= E_{2^{n-1}}$ would lead to identical conclusions.
We are about to see that the long-range nature of the Coulomb interaction yields a very detrimental scaling of the noise-induced error. This stems from a very simple fact:
when one puts together $N$ electrons with the $N$ positive charges of the nuclei, the resulting energy scales as $N^2$ if the electrons are \emph{not} allowed to screen the nuclei. 
The ground-state energy of a molecule scales as $E_0\propto N$ because the electrons properly screen the nuclei. Likewise, any reasonable variational ansatz is extensive, namely $E_V \propto N$.
However, recall that we have just seen that since the hardware noise is totally unaware of the molecule one wants to study, it will generically populate the high-energy states of $H$. 
These high-energy states, on the other hand, have a macroscopic ($\propto N$) electric dipole (classical charging energy of a capacitor).
Any noise model that creates a macroscopic electric dipole (i.e. whenever $\rho_\text{noise}$ has a finite $N$-independent weight on those states) will lead to a quadratic contribution to the observed energy (i.e. $E_{2^{n-1}}\propto N^2$). 
For instance, the relaxation channel favors a single electronic configuration (say all qubits in state zero) that is unlikely to screen the nucleus.
We therefore conjecture that the generic scaling of the noise error is
\begin{equation}
E_{\rm noise} \approx a N + b N^2,
\end{equation}
where the magnitude of the constants $a$ and $b$ will depend on the details of the problem and of the noise model. $a$ and $b$ are large energy scales, of the same order as the matrix elements $h_{ij}$ and $v_{ijkl}$ (from Hartrees to tens of Hartrees, depending on the basis set).

\subsubsection{Example of the depolarizing noise}

To corroborate this scaling, we have computed $E_{\rm noise}$ in the case of the depolarizing noise for the first $26$ atoms of the periodic table using the  PySCF package \cite{Sun2020}. 
Figure~\ref{fig:atoms} shows the energy per electron $(E_\infty-E_{\rm HF})/N$ (at this scale $E_{\rm HF}$ is indistinguishable from $E_0$) versus $N$.
It contains three important messages.
(i) First, the scale: as soon as one steps away from the minimum STO-3G basis set (which is totally insufficient to achieve chemical accuracy in any setting), the scale of the error is very large, of the order of $10$ Ha.
To put it into perspective, the right axis shows the energies in Kelvins: one quickly arrives at core sun level of temperature.
(ii) Second, the scale quickly increases when one increases the basis set from single zeta to double and triple. It means that when one improves the basis set to get a better accuracy, the noise induced error actually gets worse.
(iii) Third, we clearly observe that the noise energy per electron is not constant; 
it increases with $N$ at least linearly. This confirms the presence of the quadratic contribution $\propto N^2$ argued earlier.

\begin{figure*}[htbp]
	\includegraphics[height=5cm]{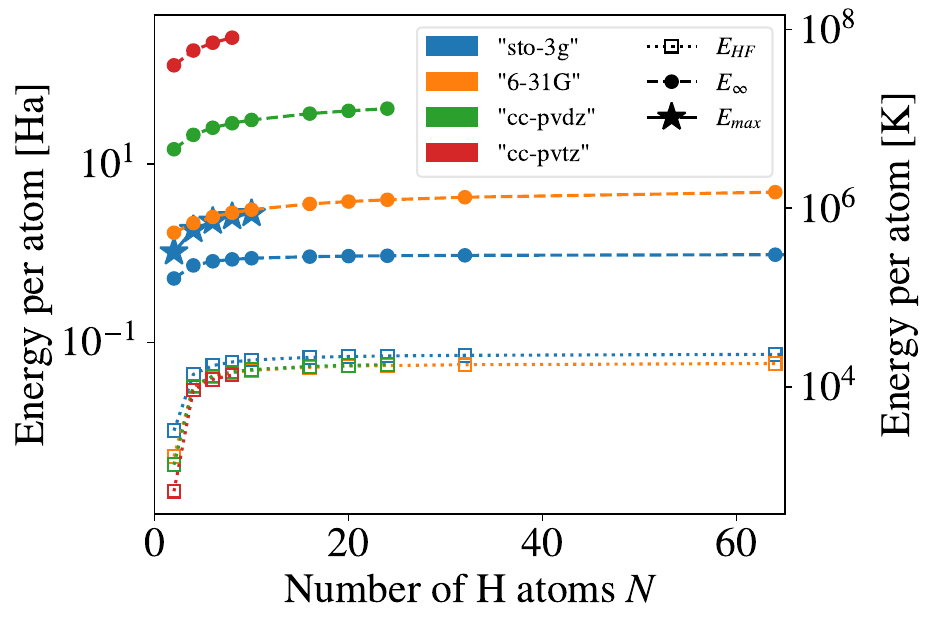}
	\includegraphics[height=5.2cm]{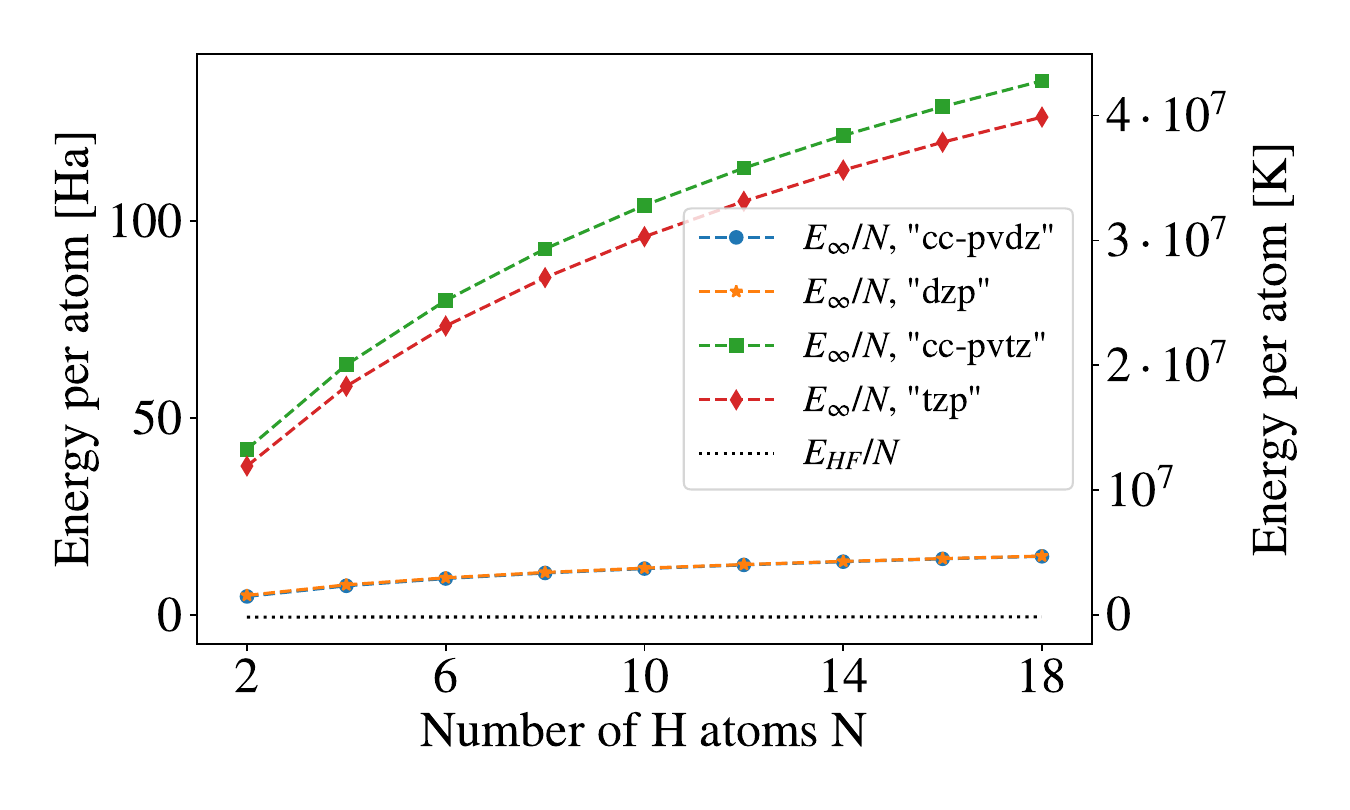}
	\caption{Different energy scales of a chain of $N$ hydrogen atoms. Left panel: symmetric setting: all atoms are treated with the same basis set. Shown are the Hartree-Fock energy $E_{HF}$ and infinite temperature energy $E_\infty$ per atom, for the "STO-3G" ($2$ qubits per atom), "6-31G" ($4$ qubits per atom) ~\cite{Ditchfield1971}, "cc-pVDZ" ($10$ qubits per atom) and "cc-pVTZ" basis ($28$ qubits per atom) ~\cite{Dunning1989}. For the "STO-3G" basis we have also plotted the highest excited state energy $E_{max}$ up to $N=10$ (regime where exact diagonalization is possible).
In all curves, the energy per atom is offset by its Hartree-Fock value for $N=2$ in the
		 "STO-3G" basis.
(Right) Effect of an assymetry on the large energy spectrum. 
For the rightmost-half of the chain we use the 'sto-3g' basis while for the leftmost-half we use more precise basis sets: "cc-pvdz", "dzp", "cc-pvtz", "tzp". 
This calculation shows that any asymetry, originating from the presence of different atoms or here
simply on the different treatment of the same atom, leads to the formation of a large electric dipole in the energy spectrum.
\label{fig:Escale_hchain}
	}
\end{figure*}

To understand better the behavior of $E_\infty$, we perform another study:  
a linear chain of $N$ hydrogen atoms (a popular benchmark).  
The left panel of Figure~\ref{fig:Escale_hchain} shows the energy $E_\infty/N$ versus $N$ in different basis sets, calculated with the PySCF package \cite{Sun2020}.
The figure confirms the above points (i) and (ii). 
Also shown is the Hartree-Fock energy (empty symbols, almost basis set independent at this scale) and the maximum energy of the Hamiltonian ($E_{2^n}$, star, only for STO-3G) which is, as expected, of the same order of magnitude.
However in this figure, point (iii) is not visible: 
after an initial rise, $E_\infty/N$ actually saturates, there is no building up of a macroscopic dipole. This is due to the high symmetry
of the problem: except on the two ends, all the atoms are essentially equivalent, hence the electrons are distributed evenly by this noise model. In real applications, however, this is seldom the case as molecules are made of different atoms with
different environments. To showcase a more generic case, we have modified the basis set of the hydrogen atom chain in the following way: the first $N/2$ atoms are treated with a small basis set (STO-3G) while the last $N/2$ atoms are treated with a larger basis set.
This, arguably artificial, setting should be understood as a pedagogical extreme limit.
The results are shown in the right panel of Figure~\ref{fig:Escale_hchain}.
We observe a strong $N$ dependence (iii) with $E_\infty$ at least ten times larger than in the symmetric setting.
In the simple error model considered here, ($\rho_{\rm noise} \propto I_{\rm d}$), the noise
essentially populates the available orbitals in a uniform way. The point we are trying to make is that such a population is likely \emph{not to} properly screen the nuclei charge, and therefore to have a dire cost in energy. This is shown in a particularly striking way on the right panel, where the asymmetry in the basis set leads to the advent of a macroscopic dipole [and the energy increases quadratically with $N$: point (iii)]. Even for the case on the left, which is highly symmetric
(all atoms play an identical role except for the two on the edges of the chain), the different orbitals are not identical, so that increasing the basis set is very costly in energy [point (ii)]. In anycase, the most important information from these curves is the scale of the noise energy: of the order of $10$ Ha, i.e. four orders of magnitude larger than the targeted accuracy [point (i)].

\subsubsection{Choice of ansatz}
A crucial piece of the VQE algorithm is the choice of the variational ansatz
used in the calculation: one aims at a quantum circuit as shallow as possible, yet
the ansatz must be sufficiently expressive to capture the ground state of the molecule
and also permit the associated optimization. Of interest to us here is the scaling of the
number of gates $N_g$ of the ansatz with the size $N$ of the molecule. A lower bound for $N_g$
is actually the number of free parameters of the ansatz since one gate can encode at most $O(1)$
parameters (in practice there are often many more gates than parameters, this is really a lower bound).

Following the difficulties encountered with the "hardware efficient" ansatz (with respect in particular to the presence of barren plateaus), the community started to consider ans\"atze inspired by successful approaches in computational chemistry such as the UCC ansatz. These ans\"atze have relatively stiff scalings (e.g., $N^6$ for UCCSDT with single double and triple excitations, not taking into account overheads stemming from fermion-to-qubit encodings or energy measurements). However these are the scalings that are also needed in {\it classical} coupled cluster calculations, so it is very possible that this scaling is already close to optimal.
Recently, some authors have attempted to use simpler ans\"atze.
For instance, ADAPT-VQE \cite{Grimsley2019} is inspired by a trotterized version of UCC and constructs the ansatz iteratively by adding the operators one by one from a pool of possible excitations.
Due to the trotterization (often found in UCC-VQE too), the ansatz is somewhat closer to configuration interaction than coupled cluster and as such it is not clear how it performs with respect to size consistency. 
A significant reduction up to 60\% was observed in the number of parameters used with respect to UCC with a better precision, at the cost of extra sampling.
Other algorithms try to avoid the need for the optimization part (and associated gradient calculation) of VQE and try to find schemes that converge naturally to the ground state.
One such construction approaches the ground state adiabatically through a randomized algorithm \cite{Granet2024} that must be averaged over different circuits with random insertions.
One could also construct a Lindbladian dynamics designed to converge to the ground state \cite{Ding2024}, use auxiliary qubits as a heat sink to cool down the system to the ground state
\cite{Polla2021}, and engineer a transformed
Hamiltonian \cite{Motlagh2024} designed for cooling. Similar techniques can be used to compute Green's functions in impurity models \cite{Bertrand2024}.
These techniques have mostly been demonstrated on very small scales. More research will be needed
to study their potential for providing an advantage both at the practical and 
even theoretical level \cite{Gharibian2023}. 

In any case, irrespectively of the type of ansatz and scheme used to optimize it towards the ground state, it is difficult to imagine that the scaling of $N_g$ would be more favourable than $M^4$.
Indeed, one needs at the very least as many gates (hence angles) as the number of parameters that define the problem itself (the $M^4$ Coulomb integrals in \eqref{eq:hamiltonian}). 
What is more, such an optimum situation assumes that one has a perfect parametrization of the corresponding $M^4$-dimensional manifold inside the much larger $2^M$-dimensional Hilbert space, a very unlikely situation.
Note a recent proposal~\cite{Luo2025} for approximate simulations of the dynamics where the scaling is reduced from $M^4$ for a Trotter simulation to $M^2$ at the cost of introducing extra qubits to rewrite the Coulomb tensor in a diagonal form. But whether this work can be extended to change the scaling of VQE ansatze remains unclear.

\subsubsection{Application: criterion estimation for Benzene}

We are now in position to answer the question:
what does the criterion Eq.~\eqref{eq:vqe_criterion} imply in practice?
It is generally accepted that a quantum chemistry calculation must reach  "chemical accuracy",  ${\eta_{\rm chem} = 1 {\rm kcal/mol}\ \approx 1.6 {\rm mHa} \approx 500 K}$ (for energy differences).
As we have seen, a very conservative estimate for the noise energy is $E_{\rm noise} \ge 1 {\rm Ha}$, likely much more.
For the $H_2$ molecule in a minimum basis set of just two orbitals per atom (STO-3G), one gets ${E_\infty - E_0 = 1.02 {\rm Ha}}$ for the depolarizing noise, so that we estimate ${E_{noise} - E_V}$ to be of the order of $1$Ha in this simple case. 
On the other hand, one needs a variational ansatz expressive enough to reach $\eta_{\rm chem}$, which provides a constraint $N_g\ge N_g(\eta_{\rm chem},N)$.

To be concrete, we consider a recent blind test benchmark on the benzene molecule \cite{Eriksen2020}.
Benzene is a nontrivial calculation for classical approaches.
Yet \cite{Eriksen2020} showcased that a variety of classical techniques arrived at chemical precision using $30$ electrons distributed on $108$ orbitals. It can be considered as a minimum target for VQE:
anything smaller is likely to be addressable by classical means.
Using the UCC ansatz, inspired by the successful coupled cluster approach used in quantum chemistry, would require to include at least single, double and triple excitations (actually quadruple would probably be needed too), which translates into $N_g> N^6$ gates.
This is probably a very conservative estimate that ignores the additional cost to implement the
(non-local) fermionic operators with qubits as well as many prefactors.
One arrives at a noise level $\epsilon \le 10^{-3}/(30)^6$, which translates into $\epsilon \le 10^{-12}$, that is \emph{many} orders of magnitude below the best existing quantum hardware. Even if error mitigation could be maintained at a level of $G=0.001$ (which is doubtful for such a large system, see the discussion in the next paragraph), the corresponding bare error level needed would remain extremely small 
$\epsilon_{\rm bare} \le 10^{-9}$ for a problem that is already within reach of several classical techniques.

Such a low level of noise is difficult to imagine without a full-fledged implementation of quantum error correction.
Since quantum error correction itself comes with an important overhead in term of computational time, one wants to avoid the accumulation of statistics required in VQE. We will thus turn to a different algorithm, QPE, in Sec.~\ref{sec:qpe}.

\subsubsection{Prospects of error mitigation}
\label{sec:error_mitig}
As we have argued in the above sections, decoherence introduces a large bias to the sought-after energy and is therefore very harmful.
The reason for this bias being large is twofold: noise generically produces high-energy states, and is a multiplicative process (hence grows exponentially quickly). In this paragraph, we quickly discuss the possibility to use error mitigation techniques to alleviate this problem. We emphasize that the noise estimates at which we have arrived are agnostic to the method used to obtain them: They are estimates of the effective noise levels per gate needed to be achieved, possibly after error mitigation or even error correction, not necessarily estimates of the bare error level on the physical qubits. 

Error mitigation techniques have become increasingly popular over the last few years \cite{Li2016a,Temme2017, Endo2017,Endo2021,Ferracin2022, Cai2022, Lolur2022, Brien2022, vandenBerg2023}.
These techniques, including probabilistic error cancellation, zero-noise extrapolation, assignment error mitigation, etc., essentially consist in reducing the bias on the energy estimator using more or less precise information on the noise processes at play.
This comes at a steep price: the variance of the energy estimator increases exponentially with the number of gates of the circuit, and even, in some cases, the number of qubits \cite{Takagi2021, Tsubouchi2023, Quek2023}.
This in turn, leads to a number of required shots, and thus a run time, that increase exponentially.

Let us be more specific with the concrete example of probabilistic error cancellation \cite{Temme2017, Endo2017}. In this method, to suppress the bias in energy in a noisy
VQE, caused by the deviation between the noisy energy $E_{\mathrm{noise}}=\mathrm{Tr}(\rho_{\mathrm{noise}}H)$
and the perfect energy $E_V=\langle\Psi_{V}|H|\Psi_{V}\rangle$,
one decomposes the ``perfect'' channels $\mathcal{E}_{\mathrm{perfect}}^{(k)}(\rho) = U^{(k)} \rho (U^{(k)})^\dagger$ (with $U^{(k)}$ the $k$th gate in the variational circuit)
as a linear combination of the actual (noisy) channels $\mathcal{E}_l(\rho)$
that are realized in the quantum computer:
\begin{equation}
\mathcal{E}_{\mathrm{perfect}}^{(k)}=\sum_{l}q_{l}^{(k)}\mathcal{E}_{l},\label{eq:lc_channels_pec}
\end{equation}
with the underlying assumption that these noisy channels form an independent
family, and the $\sum_{l}q_{l}^{(k)}=1$ to preserve the trace-preserving
character (but the $q_{l}^{(k)}$ may be negative). The perfect energy
can thus be decomposed as a very large sum, which is sampled through
Monte Carlo:
\begin{widetext}
\begin{align}
E_{\mathrm{perfect}} & =\sum_{l_{1}\dots l_{N_{\mathrm{g}}}}q_{l_{1}}^{(1)}\cdots q_{l_{N_{\mathrm{g}}}}^{(N_{\mathrm{g}})}\mathrm{Tr}\left(H\mathcal{E}_{l_{N_{\mathrm{g}}}}\circ\mathcal{E}_{l_{N_{\mathrm{g}}-1}}\circ\cdots\mathcal{E}_{l_{1}}(\rho_{\mathrm{i}})\right)\nonumber \\
 & =\Gamma\sum_{l_{1}\dots l_{N_{\mathrm{g}}}}p_{l_{1}}^{(1)}\cdots p_{l_{N_{\mathrm{g}}}}^{(N_{\mathrm{g}})}\mathrm{s}(l_{1},\dots l_{N_{\mathrm{g}}})\mathrm{Tr}\left(H\mathcal{E}_{l_{N_{\mathrm{g}}}}\circ\mathcal{E}_{l_{N_{\mathrm{g}}-1}}\circ\cdots\mathcal{E}_{l_{1}}(\rho_{\mathrm{i}})\right)\\
 & \approx\frac{\Gamma}{N_{s}}\sum_{i=1}^{N_{s}}\mathrm{s}(l_{1}^{(i)},\dots l_{N_{\mathrm{g}}}^{(i)})\mathrm{Tr}\left(H\mathcal{E}_{l_{N_{\mathrm{g}}}^{(i)}}\circ\mathcal{E}_{l_{N_{\mathrm{g}}-1}^{(i)}}\circ\cdots\mathcal{E}_{l_{1}^{(i)}}(\rho_{\mathrm{i}})\right).\label{eq:pec_formula}
\end{align}
\end{widetext}
Here, the probabilities $p_{l}^{(k)}=q_{l}^{(k)}/\sum_{l'}q_{l'}^{(k)}$
are introduced to deal with the possible negativity of the $q_{l}^{(k)}$'s,
$\Gamma=\prod_{k=1}^{N_{g}}\sum_{l'}\left|q_{l'}^{(k)}\right|$ and
$\mathrm{s}(l_{1},\dots l_{N_{\mathrm{g}}})=\prod_{k=1}^{N_{g}}\mathrm{sign}(q_{l_{k}}^{(k)})$.
The advantage of the summand of Eq. (\ref{eq:pec_formula}) is that
it can be estimated with the noisy computer at hand by running a circuit
with (noisy) gates $\mathcal{E}_{l_{1}^{(i)}},\dots,\mathcal{E}_{l_{N_{\mathrm{g}}}^{(i)}}$
and measuring $H$ at the end, resulting in an unbiased estimate of
$E_{\mathrm{perfect}}$.
The price to pay is, however, twofold.
First, one
needs to know precisely the noise models of the hardware to be able
to perform the decomposition of Eq. (\ref{eq:lc_channels_pec}). This will require one
to perform expensive tomography experiments, the cost of which will increase algebraically as one requires more precision or exponentially as one includes more complex noise processes with more qubits (such as crosstalk between different gates). There are also underlying assumptions such as the absence of drift (due to e.g. the two-level systems that are currently a very important limiting factor of superconducting qubits) or the precision of the tomography that ultimately limits the precision reachable by this approach. 
Second, and more fundamentally,
the estimator comes with a statistical uncertainty  $\Delta E_{\mathrm{shot}}$ that scales with
the $\Gamma$ factor, which can be rewritten as $\Gamma=\prod_{k=1}^{N_{g}}\left(1+2\eta_{k}\right)$,
with the so-called ``negativity'' $\eta_{k}=\sum_{l,q_{l}<0}q_{l}^{(k)}$.
To get an intuition how it scales, let us suppose that all gates have
the same negativity $\eta$. Then
\begin{equation}
\Gamma\approx e^{2\eta N_{\mathrm{g}}}.\label{eq:negativity_exponential}
\end{equation}

Let us make a connection to our error threshold $\epsilon_\mathrm{max}$ (Eq.~\eqref{eq:vqe_criterion}). So far, our main concern was the bias $\Delta E$ in the energy, $\Delta E=\left(1-F\right)(E_{\mathrm{noise}}-E_{V})\approx\epsilon N_{g}(E_{\mathrm{noise}}-E_{V})$. For a finite number of shots, however, the total energy error has a contribution from this bias and from statistical shot noise:
\begin{equation}
\Delta E_{\mathrm{tot}}=\sqrt{\Delta E^{2}+\Delta E_{\mathrm{shot}}^{2}}.
\end{equation}
With probabilistic error cancellation, we can a priori send $\Delta E$ to zero, but $\Delta E_{\mathrm{shot}}$ is multiplied by $e^{2\eta N_{g}}$. To keep $\Delta E_{\mathrm{shot}}$ constant, we thus need to increase $N_{\mathrm{s}}$ by a factor $e^{4\eta N_{g}}$.

Note that this exponential increase in the complexity is rather general: the exponential loss of fidelity can be viewed as a loss of information on the system. To fight it, one must acquire extra information either during the performance of the circuit (that would be
quantum error correction with the syndrome measurements) or after (this is mitigation). In the latter case, the loss of information has had the time to accumulate exponentially and must be matched by some sorts of extra measurements. This argument applies irrespectively of the type of mitigation that is being used. The noise can only be mitigated to the extent that it is well understood. For instance, mitigating crosstalk proves to be much harder than simple one-qubit gate errors \cite{Perrin2024}.

New classes of methods such as Refs.~\cite{Robledo-Moreno2024,Yu2025} are delegating a much larger share of the calculation to classical supercomputers. In Refs.~\cite{Robledo-Moreno2024,Yu2025} the role of the quantum computer is much narrower than in a VQE calculation: it merely proposes the relevant Slater determinants that will be included in a subsequent (purely classical) configuration-interaction type of calculation. Establishing the advantage of such techniques will require more studies. In particular, they face the same limitation as the classical techniques they are based on, so they can at best provide a limited improvement compared to those techniques. Should they provide a genuine improvement, it would also be interesting to study if the proposal step could be done classically.

We therefore think it is unlikely that error mitigation will solve the noise problem discussed above: in practice, the number of shots cannot be scaled exponentially with system size, leading to a residual bias.
In any case, the above criterion still holds in the presence of practical error mitigation:
One just needs to replace $\epsilon$ by $\epsilon^\mathrm{\rm mitigated}$ in Eq.~\eqref{eq:vqe_criterion}.
For small molecules with less than $15$ orbitals, it has been estimated~\cite{Dalton2024} that error mitigation might improve by one or two orders of magnitude the maximum allowed error per gate for VQE (see also \cite{Brien2022} for a 20-qubit case). For larger molecules, because of the scaling problems mentioned above, the gain one can expect from mitigation should be even lower, leaving the threshold of $10^{-12}$ for Benzene out of reach.

\subsection{VQE versus its classical competitor, the variational Monte Carlo (VMC) algorithm}

\begin{table*}[t!]
\setlength{\tabcolsep}{.7em}
\begin{tabular}{ |p{5cm}||p{4.5cm}|p{4.5cm}|}
\hline
 Problem & VQE (quantum) & VMC (classical) \\
 \hline
 (I) Statistical uncertainty & Slow $\sigma/\sqrt{N_s}$ convergence  with a fixed pre-factor $\sigma$ & Also $\sigma/\sqrt{N_s}$ but $\sigma$ vanishes as the ansatz approaches the ground state \\
 \hline
 (II) Optimization of the wave function &  Gradient estimation very costly $O(N_{\vec\theta})$  & Gradient calculation is cheap  $O(1)$ (full gradient for the cost of one evaluation) \\
    & Possible barren plateau & Possible barren plateau \\
\hline
 (III) Hardware availability  & Need "fault-tolerant quantum computer" level of precision (this work)   & Works on existing CPU, GPU and TPU. Trivial parallelism. \\
 \hline\hline
 Sampling & direct   & Markov chain (usually) or direct (e.g. autoregressive neural networks) \\
 \hline
\end{tabular}
\caption{Comparison between VQE and VMC for the three main difficulties faced by these algorithms. 
\label{vqe_vs_vmc}}
\end{table*}

In the previous section, we discussed a necessary condition for VQE to be able to reach chemical accuracy.
Here, we stress the fact that a condition for quantum advantage is more stringent: VQE must be faster than competing purely classical algorithms.
There are essentially three classes of classical algorithms that we could think of. 
(i) The first class corresponds to algorithms designed to simulate quantum circuits such as in Refs. \cite{Zhou2020,Ayral2023,Angrisani2024}. Trivially, if the quantum
circuit of the VQE ansatz can be simulated, there is not much point in building the corresponding quantum computer.
(ii) Second are
state-of-the-art computational chemistry techniques. They involve a variety of approaches
depending on the size of the molecules being studied, the accuracy needed and the type
of constituents (e.g. organic molecules with close shells versus the presence of transition metals that imply the need for multi-reference calculations). These techniques include density functional theory, coupled cluster methods, various variants of configuration interaction, quantum Monte Carlo and the density matrix renormalisation group (DMRG) tensor network approach. All these techniques beat VQE working on existing quantum hardware by a large margin. The question is therefore not how they perform but how they scale when going to more complex and/or more accurate calculations.
To perform a meaningful comparison with VQE in this context, we therefore consider (iii) a third class of classical techniques:
approaches that share as many common points with VQE as possible, in particular the property of being stochastic. 

A natural comparison point for VQE is the classical algorithm of variational Monte Carlo (VMC) \cite{Ceperley1977}---an old technique that has been experiencing an important revival in the last few years thanks to the success of deep neural networks based ansatz.
Indeed, the introduction of neural network variational wave functions with a large number of parameters $N_\theta$ has led to very encouraging results~\cite{Hermann2023,Choo2020}.
In VMC, one also considers a variational wavefunction $\ket{\Psi_V(\vec \theta)}$.
This function can be anything as long as one is able to compute its value $\langle \vec s \ket{\Psi_V}$ for a given configuration $\ket{\vec s}$.
This includes wavefunctions explicitly written in term of a neural network as well as physically-motivated wavefunctions, or combinations thereof. However, these ans\"atze are typically \emph{not} written in terms of a quantum circuit in order to avoid the large computational overhead associated with the contraction of the associated tensor network.
In VMC, the variational energy is expressed as
\begin{eqnarray}
E_V = \sum_{\vec s} E_{\vec s} \  | \langle \vec s \ket{\Psi_V}|^2, \label{eq:VMC}
\end{eqnarray}
with the so-called local energy 
\begin{equation}
E_{\vec s} =\frac{\bra{\Psi_V} H \ket{\vec s}}{\langle \vec s \ket{\Psi_V}} =\sum_{\vec s_1}  \bra{\Psi_V} \vec s_1\rangle \bra{\vec s_1} H \ket{\vec s} \frac{1}{\langle \vec s \ket{\Psi_V}}.
\end{equation} 
The sum over configurations in~\eqref{eq:VMC} is sampled by a Monte Carlo algorithm, namely one draws $N_s$ samples from the distribution $|\langle \vec s \ket{\Psi_V}|^2$ and estimates the energy $E_V$ as
\begin{equation}
E_V \approx \frac{1}{N_s} \sum_{i=1}^{N_s} E_{\vec s_i}.
\end{equation}

Both VQE and VMC are based on iteratively adjusting a complex variational state $\ket{\Psi_V}$. Yet, they have important differences:

(1) as we just saw, VQE suffers from a steep loss in quality as the circuit depth increases, imposing a strong limitation on the number of gates, and thus the expressiveness, of the quantum circuits that can be reliably executed. By contrast, VMC does not suffer from decoherence. 

(2) The sampling method is very different. Quantum computers are ideal in that respect, they provide direct sampling.
VMC often needs to use a Markov chain method such as the Metropolis-Hastings algorithm, which introduces an overhead.
However, some modern wavefunctions, like the restricted Bolzmann machine or those constructed from 
autoregressive neural networks such as transformers or recurrent neural networks, do provide direct sampling. 

(3) In both cases, the statistical error decays as $\sigma/\sqrt{N_s}$. However, the prefactors $\sigma$ behave in very different ways. 
In VMC $\sigma$ is given by the standard deviation of the local energy $E_{\vec s}$.
In the extreme case where the variational wavefunction coincides with the ground state, $E_{\vec s} = E_0 \ \forall~ \vec s$: the standard deviation $\sigma$ vanishes entirely, i.e. a single sample is enough to have infinite precision. 
This phenomenon, dubbed variance reduction, is absent in VQE.
Assuming, for simplicity, that each term $H_\alpha$ contains only one Pauli term ( $H_\alpha = \lambda_\alpha P_\alpha$), we have that $\sigma =\sqrt{\sum_\alpha \lambda_\alpha^2 / \nu_\alpha \sigma(P_\alpha)^2} $, where $\nu_\alpha$ is the fraction of the $N_s$ shots allocated to term $\alpha$, and $\sigma(P_\alpha)$ is the standard deviation of the Pauli observable $\langle P_{\alpha} \rangle$:
\begin{equation}
\sigma(P_\alpha) = \left[ \bra{\Psi_V} P_{\alpha}^2 \ket{\Psi_V} - \bra{\Psi_V} P_{\alpha} \ket{\Psi_V}^2 \right]^{1/2}.
\end{equation} 
Here, even if $\ket{\Psi_V}$ happens to be the exact ground state, it is not an eigenstate of the individual $P_\alpha$ operators and the variance thus does not vanish.
In fact, since those terms originally stem from the kinetic and Coulomb repulsion matrix elements, they have quite large magnitude (typically in the $1$ Ha range).
This absence of variance reduction in VQE reflects the fact that it is much more advantageous to measure the full Hamiltonian at once than its different components separately.
In practice, as the optimization proceeds in VMC, one wants to measure $E_V$ with more and more digits of precision.
Since $\ket{\Psi_V}$ gets closer to $\ket{\Psi_0}$, in VMC the energy variance decreases and hence the precision of the calculation also becomes better for the same experimental time (same number of samples).
In contrast, in VQE the variance of the different energies $E_\alpha$ does not decrease, hence one must increase the number of samples by a factor 100 for each new digit in precision on the measure of $E_V$.
This might constitute another important advantage in favor of VMC.

(4) In VQE and VMC, being able to optimize the energy strongly relies on the ability to calculate the gradient $\partial E_V/\partial\vec\theta$.
Classical techniques use automatic differentiation and in particular, in the context of neural networks, the backpropagation algorithm.
Very notably, the backpropagation algorithm provides the entire gradient (i.e. for all parameters at once) at essentially the same cost as the evaluation of the wavefunction itself.
In contrast, quantum computers typically need to perform many more ($O(N_{\vec\theta})$) computations \cite{Abbas2023}. Given that useful variational ans\"atze typically have thousands or hundreds of thousands parameters, this could constitute an important penalty.

(5) The two techniques have access to different sorts of ans\"atze. VMC is more versatile in that respect and can simulate VQE ans\"atze (as long as the quantum circuit depth is not too deep) while the reverse is not true. For instance, VMC can use Tensor network states that give an efficient representation of weakly entangled states. But VMC can also use neural network states which can sustain highly entangled states including volume law entanglement. In VQE, the quantum nature of the hardware allows, in principle, to build arbitrarily entangled states, but in practice the depth might be excessive. This is particularly true for the existing (NISQ) hardware which is limited to shallow circuits, hence limited entanglement. At the moment it is unclear if one technique has an advantage in that respect. One should keep in mind that the expressivity of the ansatz is strongly linked to the ease/difficulty to perform the optimization. An ansatz which is too expressive
(for instance the so-called "hardware efficient ansatz" in VQE) leads to barren plateaus as most of the parameter space corresponds to almost random (chaotic) states (see the above discussion on barren plateaus).

The conclusions of this comparison between VQE and VMC are summarized in Table~\ref{vqe_vs_vmc}.
What one should take away from this exercise is that, in contrast to e.g. Shor's algorithm, which is expected to speed up the factorization
of large numbers exponentially, there are no similar indications for VQE, quite the opposite actually.
Even if we had perfect quantum computers, VMC might still remain superior to VQE. (See also \cite{Mazzola2023} for a discussion including other classical Monte Carlo methods).

\section{A criterion for quantum phase estimation (QPE): estimating the orthogonality catastrophe}\label{sec:qpe}

\begin{figure*}
	\includegraphics[width=17.5cm]{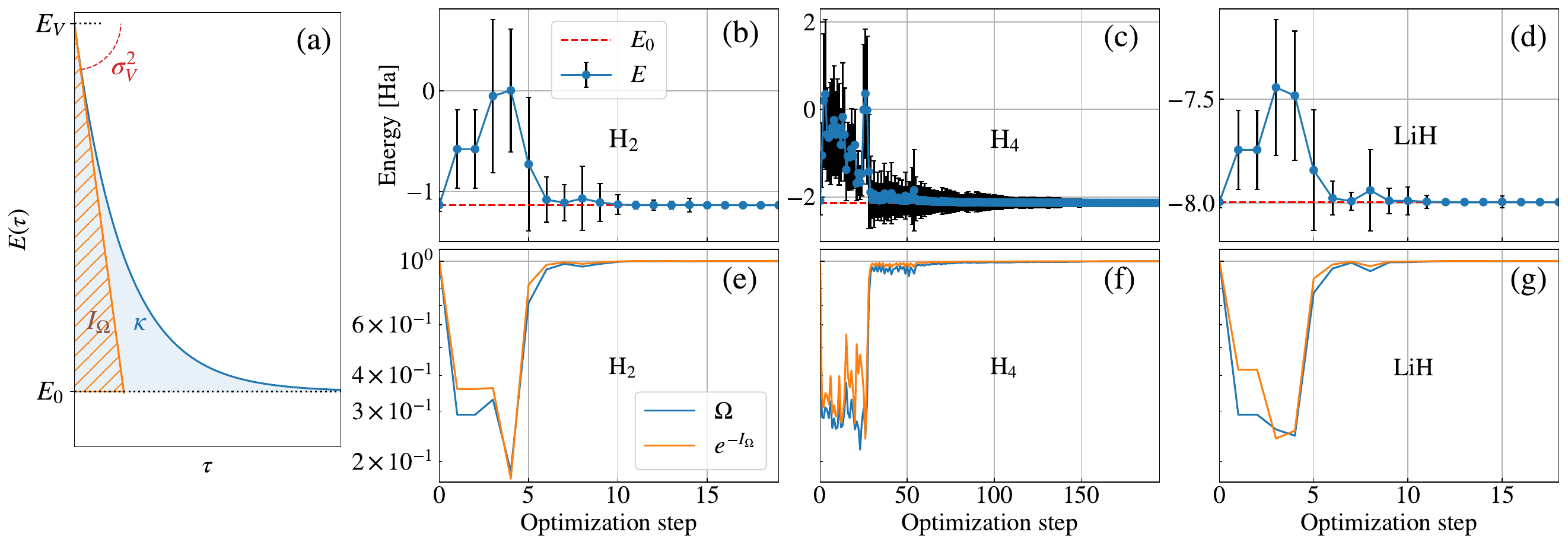}
	\caption{(a) Sketch of the energy $E(\tau)$ vs imaginary time. 
	 The area $ \kappa$ under the energy curve directly provides the overlap $\Omega=e^{-\kappa}$. We approximate $\kappa$ with the more easily accessible orange-shaded area $I_\Omega$. [(b) - (d)]: Energy $E_V$ vs optimization step in VQE simulation, the "error bar" corresponds to the standard deviation $\sigma_V$. 
	 [](e) - (g)]: Overlap $\Omega$ and $e^{-I_\Omega}$ vs optimization step. \label{fig:varindex}
	}
\end{figure*}

\begin{figure}
	\includegraphics[width=8.5cm]{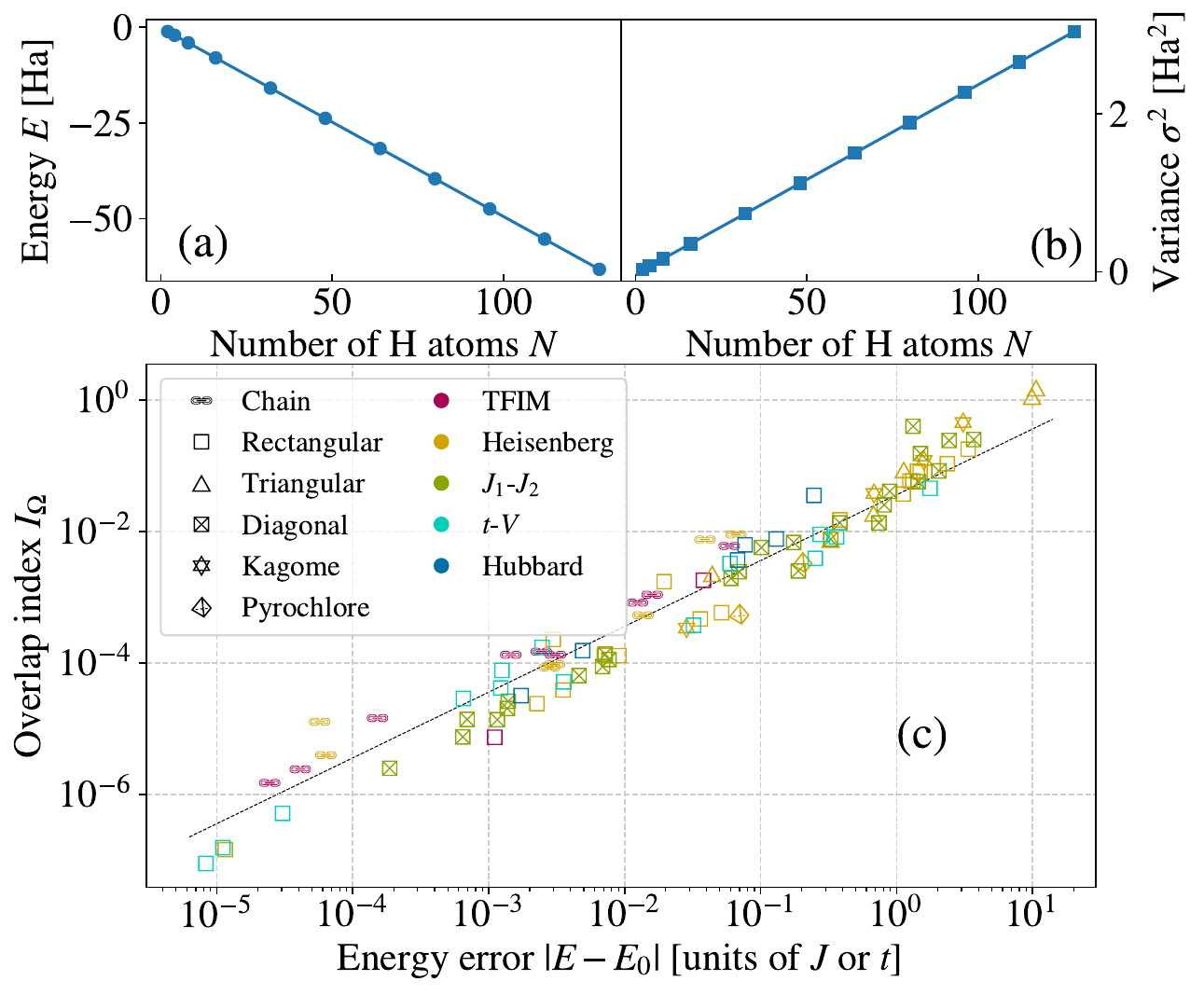}
	\caption{\label{fig:scaling} 
	 Scaling of the Hartree-Fock energy (a) and the energy variance (b) vs number of atoms in a hydrogen chain. 
	(c): Overlap index $I_\Omega$ versus energy error $|E-E_0|$ for the variational ansatz of the \cite{Wu2023} data set.  The dashed line is a linear fit
	 $I_\Omega \approx 27.8 |E-E_0|$. }
\end{figure}

We now turn to the second criterion, relevant for the quantum phase estimation (QPE) algorithm. 
Our goal is to quantify the orthogonality catastrophe phenomenon  \cite{Anderson1967}, which has been proposed recently as a possible strong limitation to the applicability of QPE \cite{Lee2022}.
Our criterion provides a direct measure of this phenomenon. It can be computed in situations where one does not have access to the full solution $| \Psi_0 \rangle$ of the problem, namely the actual cases that would be targeted by QPE.

\subsection{Role of the overlap in QPE}

QPE starts from a guess input state $\ket{\Phi}$ and extracts one of the eigenenergies $E_i$ of $H$ with a probability $p_i = |\langle \Phi  \ket {\Psi_i} |^2$ while projecting the input state to the corresponding eigenstate $\ket{\Psi_i}$ of $H$.
In general, one designs $\ket{\Phi}$ to have the largest overlap with the ground eigenstate $\ket{\Psi_0}$ of $H$, so that we will henceforth set $i=0$, and call
\begin{equation}
\Omega= |\langle \Phi | \Psi_0 \rangle |^2 \label{eq:overlap}
\end{equation}
the overlap of $\ket{\Phi}$ with the ground state.
$\Omega$ is thus also the success probability of QPE.
In practice, it means that QPE needs to be repeated $O(1/\Omega)$ times to find the ground state energy with high probability (or possibly $O(1/\Omega^{1/2})$ using optimal algorithms \cite{Lin2020}).

The main appeal of QPE is that it reaches a level of precision $\eta$ with a circuit depth, and thus a run time, scaling like $1/\eta$.
This is a quadratically better run time compared to VQE's $O(1/\eta^2)$ run time. 
This fundamental difference stems from the fact that the energy estimation is done in a purely quantum manner, as opposed to the classical average of energy measurements that underlies VQE.
The price to pay is a much larger circuit depth, which makes QPE appropriate only for hypothetical \cite{Waintal2019} noiseless (fault-tolerant) quantum computers.

Thus, a key necessary condition for the success of QPE is a large overlap $\Omega \sim 1$ of the input state to the ground state.
For QPE to be of practical use, it is therefore necessary to provide the QPE user with a practical means to quantify the overlap in realistic situations where the solution is not known a priori.
This is all the more crucial as the state preparation may become exponentially difficult for large systems, i.e. in the very region where QPE is expected to show an advantage over classical or NISQ methods.
Indeed, in condensed matter, the overlap between two states, however close they are at the single-body level, is believed to decrease exponentially with system size; this is the orthogonality catastrophe \cite{Anderson1967}.
This phenomenon holds even for states that share very similar energies: two states can be infinitesimally close in energy but physically very distinct.
This means in particular that there is no guarantee that a variational state prepared with an algorithm such as VQE will be a good candidate for QPE.
For instance, the difference of energy per particle between superconducting aluminum and normal-metal aluminum is extremely tiny $\sim (\Delta/E_F)^2 \approx 10^{-8}$ (with $\Delta$ the superconducting gap and $E_F$ the Fermi energy), yet the two states behave drastically differently.
In quantum chemistry, the situation has been studied in less detail: \cite{Tubman2018} looked at small molecules with small static correlations (good targets for classical calculations) and concluded that $\Omega$ could be kept to relatively high values. Similar conclusions were found in \cite{Goings2022} for a larger compound. In contrast, \cite{Lee2022}, which looked at more correlated molecules, gave indication of a relatively fast decay of the overlap.
We note that one should be careful with extrapolating calculations made on small molecules:
they may be misleadingly optimistic because the variational ansatz may have enough parameters to capture the entire Hilbert space (or a large fraction), in sharp contrast to real situations of interest. The overlap problem has lead some authors to recently consider some mitigation strategies that rely on heavy classical calculations to provide an ansatz with high enough fidelity. 
One possibility would be to use the results of a DMRG calculation since the corresponding matrix product states can be associated to a quantum circuits and then directly encoded in a quantum computer \cite{Berry2024}. In a different context, it had already been shown that, under some circumstances, a quantum circuit improves on a classical result (Corollary E.1. in~\cite{Chen2023}). We note, however, that in such a scheme, one remains constrained by the limitations of the classical approach so the quantum computer could at best give some extra precision for molecules that are already accessible to classical calculations. Whether this could be sufficient to bring an insufficient guess to chemical accuracy remains to be shown.

\subsection{A criterion for estimating the overlap and the orthogonality catastrophy}

Here, we show that the success probability $\Omega$ can actually be estimated in realistic situations where one \emph{does not} have access to the exact ground state.
Let us assume that the initial state $\ket{\Phi}$ fed to QPE has been obtained, for instance using a variational computation like VQE. However note that our demonstration is general, valid for any given state preparation method. 
Then, we use a theorem proved by one of us in \cite{Mora2007} to estimate the overlap $\Omega$ \eqref{eq:overlap} of this state $\ket{\Phi}$ with the ground state $\ket{\Psi_0}$.
Let us consider the wavefunction $\ket{\Psi(\tau)} = \frac{1}{\sqrt{Z}} e^{-H\tau} \ket{\Phi}$, where the factor $Z=\bra{\Phi}e^{-2 H\tau} \ket{\Phi}$ ensures normalization. 
This wavefunction appears in various techniques (e.g. Diffusion Monte Carlo or Green function Monte Carlo) that project the initial wavefunction onto the ground state using so-called imaginary-time evolution, since $\lim_{\tau\rightarrow\infty} \ket{\Psi(\tau)} = \ket{\Psi_0}$.
A typical output of these methods is the energy $E(\tau) = \bra{\Psi(\tau)} H \ket{\Psi(\tau)}$ as a function of $\tau$ as sketched in Fig.~\ref{fig:varindex} (left panel).
The overlap $\Omega$, giving the success probability of QPE, is simply related to the area $\kappa$ under this curve as
\cite{Mora2007},
\begin{align}
\Omega = \exp(-\kappa), \mathrm{~with~} \kappa = \int\limits_{0}^{\infty} d\tau (E(\tau) - E_0).
\label{eq:kappa}
\end{align}
While $\kappa$ is not necessarily easy to compute, we find that a good proxy can be obtained by considering the area $I_\Omega$ of the dashed triangle in the left panel of Fig.~\ref{fig:varindex}.
This area can be calculated from the knowledge of the initial energy $E_\Phi = E(\tau = 0)$, the energy variance of the preparation ansatz $\sigma_\Phi^2 =  - \partial_\tau E (\tau = 0)$, and an estimate (that need not be very accurate) of the ground-state energy $E_0$:
\begin{equation}
I_\Omega \equiv \frac{(E_\Phi - E_0)^2}{2\sigma_\Phi^2}. \label{eq:def_varindex}
\end{equation}
We call $I_\Omega$ the "overlap index" of the ansatz $\ket{\Phi}$.
$I_\Omega$ provides an estimate of $\Omega$ through the relation
\begin{equation}
 \Omega \approx e^{-I_\Omega}.
 \label{eq:fidapprox}
\end{equation}
While Eq.~\eqref{eq:kappa} is an exact formula, Eq.~\eqref{eq:fidapprox} is an approximation that relies on the energy $E(\tau)$ to smoothly relax to the ground state.
When $E(\tau)$ is convex (this has always been the case in previous Green's function quantum Monte Carlo calculations performed by the authors), then
Eq.~\eqref{eq:def_varindex} gives an exact upper bound $\Omega \le e^{-I_\Omega}$, which is sufficient for our purpose.
The associated success criterion for QPE is naturally $\Omega \sim 1$, i.e. $I_\Omega \ll 1$.
This criterion depends on the energy and variance of the ansatz.
It is totally independent on the imaginary-time evolution used for its derivation, and it is much easier to calculate than the actual overlap $\Omega$.
Actually, in cases where one would be able to calculate $\Omega$ directly, there would simply be no need for a quantum computer, as one would already have the entire information on the given state.

To corroborate the validity of this overlap estimation, we have performed VQE simulations for several small molecules in simple orbital bases, computing both the variational energy and the variance.
For these small systems (up to 8 qubits), the exact ground state energy in the given orbital basis can be calculated as well using exact diagonalization (aka full configuration interaction).
This allows to compute the exact overlap $\Omega$ and check that the estimate Eq.~\eqref{eq:fidapprox} holds.
On bigger systems, one would rely on estimates for $E_0$ such as those currently used in quantum chemistry where one extrapolates from a sequence of increasingly accurate calculations (e.g. CCSD, CCSDT and CCSDTQ calculations).
We use the myQLM-fermion package~\cite{Haidar2022}, a one-layer UCC ansatz and a minimum basis set (STO-$3$G for H$_2$ and H$_4$ molecules; the $6$-$31$G basis and active space selection to reduce the Hilbert space dimension from $22$ to $4$ for LiH).

The convergence of the results versus optimization steps is shown on the right panels of Figure~\ref{fig:varindex}.
Note that the "error bars" in the upper panels stand for the variance of the variational ansatz.
In the lower panels, we observe a very good match between the right and left-hand sides of Eq.~\eqref{eq:fidapprox}, which shows that the overlap index is a good proxy for $\Omega$.
Our results indicate that the joined calculation of the energy $E_V$ and the variance $\sigma_V^2=\langle H^2 \rangle - E_V^2$ makes variational calculations much more valuable.
Note that in the H$_4$ simulation, which uses $8$ qubits in contrast to H$_2$, and LiH, where the number of qubits is only $4$, the convergence is much slower.

We end with a discussion of the scaling that one may expect for $\Omega$.
A reasonable variational energy is an extensive quantity $E_V \propto N$.
Likewise, the variance is also likely to be extensive $\sigma^2 \propto N$ (this is true if the energy is roughly the sum of local terms).
It follows that the overlap index is generically an extensive quantity $I_\Omega = \alpha N$, from which one concludes that the overlap decreases exponentially 
\begin{equation}
	\Omega \approx e^{-\alpha N}.
\end{equation}	
This is the orthogonality catastrophe in the context of variational calculations.
To illustrate the above statements, the top panels of Figure~\ref{fig:scaling} show the Hartree-Fock energy (left) and variance (right) of hydrogen chains of up to $N=128$ in the STO-$3$G basis set.
Both indeed scale linearly with $N$, as advertised.

In a slightly different context, a recent work \cite{Wu2023} has aggregated a large dataset of energies and variances of variational ans\"atze of various condensed-matter systems of various sizes using various methods.
The bottom panel of Figure~\ref{fig:scaling} shows $I_\Omega$ versus $E_V-E_0$ for the data set of \cite{Wu2023}.
We find that $I_\Omega$ is well fitted by a linear law $I_\Omega = C|E-E_0|$, with $C = 27.8 \pm 0.1$.
Since $|E-E_0|$ is an extensive quantity, this implies again the exponential decay associated with the orthogonality catastrophe.

We note that this is a generic problem of phase estimation algorithms: even more recent version of the algorithm assume an input state with a large overlap (see e.g \cite{Lin2022b}, which even requires a lower bound on this overlap, and \cite{Kiss2025}, which does not).

\section{Conclusion}
To conclude, we have proposed two criteria, one for VQE (noisy hardware) and one for QPE (fault-tolerant hardware), that are easily accessible and provide necessary conditions for the possibility of doing genuinely relevant chemistry calculations on quantum hardware.
Our preliminary estimates imply that this possibility is unlikely with the approaches and technologies that are currently pursued. 
Decoherence on NISQ devices appears as a very important roadblock for VQE, not to mention shot noise and barren plateau issues. 
There are multiple proposals to circumvent the latter two problems.
One may involve a deeper hybridization between classical and quantum codes, as in \cite{Robledo-Moreno2024,Yu2025}, where the quantum computer is used only to produce the most relevant electronic configurations, or, in general, in hybrid Monte Carlo codes (see e.g \cite{Jiang2024a} for a review).
There is, however, no guarantee that a purely classical selected configuration interaction method would not yield better results.
As for decoherence, ideas to use noise to one's advantage have emerged recently \cite{Cubitt2023, Mi2023,Maciejewski2024}.

Fault-tolerant algorithms face, of course, a technological wall. First demonstrations of very small error codes have appeared \cite{Krinner2021, Chen2021, Bluvstein2023, DaSilva2024, Acharya2024}, but the size and features of these codes are still very far from the requirements of circuits needed e.g on QPE.
More worrying, the orthogonality catastrophe in QPE makes quantum avantage an exception rather than the rule \cite{Chan2024}. Alternatives are definitely needed.

These observations may also suggest that ground state estimation in chemistry may not be the most appropriate target for quantum computers. Besides the issues of quantum processors we outlined in this paper, this statement is also due to the comparatively good quality of classical state preparation methods.
By contrast, the same classical computers are easily thwarted by dynamical many-body phenomena, where sign problems or entanglement growth rapidly cause classical algorithms to reach their limits.
Quantum computers, on the other hand, offer a natural playground for dynamical evolution of quantum many-body problems, as envisioned by Feynman in the early 1980s and formalized in \cite{Lloyd1996}.
Hence quantum dynamics might be the soft target for quantum computing, even if a niche one, 
that one has been looking for.
There again, decoherence has to be reckoned with: in the absence of quantum error correction, purely quantum algorithms are bound to fail.
In the shorter term, hybrid algorithms, that combine advanced classical methods like tensor networks (see e.g \cite{Anselme-Martin2023}) or approximate mappings (see e.g \cite{Michel2023a}), might have the best chances to reach useful quantum advantage for quantum dynamics.

\section*{Acknowledgements}
We acknowledge funding from the French ANR QPEG and the ANR EPIQ. XW acknowledges funding from the PEPR EQUBITFLY, the ANR DADI, the ANR TKONDO and the CEA-FZJ French-German project AIDAS.

\bibliography{bibliography}
%

\end{document}